\documentclass{ws-procs975x65}

\begin{document}

\title{ELECTRO-WEAK MODEL WITHIN A 5-DIMENSIONAL LORENTZ GROUP THEORY}

\author{O. M. LECIAN$^{12a}$, G. MONTANI$^{123b}$}

\address{$^{1}$ICRA --- International Center for Relativistic Astrophysics.\\
$^{2}$Dipartimento di Fisica, Universit\`a di Roma ``La Sapienza'', P.le
Aldo Moro 5,\\
00185 Roma, Italy.\\
$^{3}$ENEA C.R. Frascati (Dipartimento F.P.N.), Via Enrico Fermi 45,\\
00044 Frascati, Roma, Italy.\\
$^{a}$E-mail: lecian@icra.it
$^{b}$E-mail: montani@icra.it}

\begin{abstract}
The Electroweak model will be geometrized in a 5-D Riemann-Cartan framework: $U(1)$ weak hyper-charge group will be worked out in a Klauza-Klein scheme, while $SU(2)$ weak isospin group will be identified to suitable bein projections of the contortion field. The possibility of introducing Ashtekar formalism in 5-D Klauza-Klein theories will be investigated.  
\end{abstract}


\bodymatter

\section{Lorentz gauge theory}\label{aba:sec1}
In general relativity, the expression of a generic isometric infinitesimal diffeomorphism and that of an infinitesimal local Lorentz transformation can be easily identified, i.e.
\begin{equation}
x^{\mu}\rightarrow x'^{\mu}(x)=x^{\mu}+\xi^{\mu}(x)=x^{\mu}+\epsilon^{\mu}_{\ \nu}(x)x^{\nu}=x^{\mu}+\tilde{\epsilon}^{\mu}(x);
\end{equation}
if spinor fields are introduced, the previous identification is no longer true, for spinor fileds $\psi$ transform according to a spinor representation of the Lorentz group $\psi \rightarrow S(\Lambda)\psi$, and behave like scalars under diffeomorphisms. For this reasons, local Lorentz transformations can be treated as an independent gauge group for spinors\footnote{we emphasize that under a non-isometric transformation spinors behave like scalars}, i.e. $\psi(x)\rightarrow S(\Lambda)(x)\psi(x)$ by the introduction of a gauge covariant derivative, i.e. $\partial_{\mu}\psi\rightarrow D_{\mu}\psi=(\partial_{\mu}-\frac{1}{2}\Sigma_{\bar{a}\bar{b}}A^{\bar{a}\bar{b}}_{\mu})\psi$ and of a bosonic Lagrangian density $\propto F_{\bar{a}\bar{b}}(A)F^{\bar{a}\bar{b}}(A)$ for the pertinent non-Abelian gauge field $A^{\bar{a}\bar{b}}_{\mu}$, which transforms like $C^{\bar{a}\bar{b}}_{\mu}\rightarrow S(\Lambda)(x)C^{\bar{a}\bar{b}}S^{-1}(\Lambda)(x)+S(\Lambda)(x)\partial_{\mu}S^{-1}(\Lambda)(x)$, where structure constants are defined as $[\Sigma_{\bar{a}\bar{b}},\Sigma_{\bar{c}\bar{d}}]=D^{\bar{e}\bar{f}}_{\bar{a}\bar{b}\bar{c}\bar{d}}\Sigma_{\bar{a}\bar{b}}$.\\
In a flat 4-D manifold, equipped of the set of bein vectors $e^{\bar{a}}_{\mu}$, the Lagrangian desity is made invariant under diffeomorphisms by the presence of the bein vectors, i.e. $L=-\frac{i}{2}\bar{\psi}\gamma^{\bar{a}}e^{\mu}_{\bar{a}}\partial_{\mu}\psi +h.c.$. In curved space-time, on the other hand, the request that Dirac algebra be still valid in a non-flat metric leads to the definition of a geomterical covariant derivative 
$\partial_{\mu}\psi\rightarrow D_{\mu}\psi=(\partial_{\mu}-\frac{1}{2}\Sigma{\bar{a}\bar{b}}R^{\bar{a}\bar{b}}_{\mu}-\frac{1}{2}\Sigma{\bar{a}\bar{b}}K^{\bar{a}\bar{b}}_{\mu})\psi$, where the bein projections of Ricci coefficients and of the contortion field are introduced, the affine connection of the manifold being $\Gamma^{\mu}_{\nu\rho}=\left\{
\begin{array}{c}
\mu\\
\nu\rho
\end{array}
\right\}-K_{\nu\rho}^{\mu}$ .Since Ricci coefficients are not primitive objects, and cannot be taken as gauge fields, the contortion field, which is a non-vanishing quantity even in flat space-time, can be identified to this gauge field.
Conserved quantities can be found by the identification of the conserved gauge charge with the bein projection of the conserved spin angular momentum tensor:
\begin{equation}\label{conserved}
Q^{\bar{a}\bar{b}}=\int d^{3}x J^{0 \bar{a}\bar{b}}=
M^{\bar{a}\bar{b}}= \int  d^{3}x  \pi_{r}\Sigma^{\bar{a}\bar{b}}_{rs}\psi_{s}= const.,
\end{equation}

\section{5-D model}
In a 5-D Kaluza-Klein scheme $V^{5}=V^{4}\oplus S^{1}$, the metric tensor components are
\begin{equation}\label{metric}
\begin{cases}
j_{55}=1\cr
j_{5\mu}=\tilde{g}'k\tilde{B}_{\mu}\cr
j_{\mu\nu}-(\tilde{g}'k)^{2}\tilde{B}_{\mu}\tilde{B}_{\nu}=g_{\mu\nu},\cr
\end{cases}
\end{equation}
where $k$ is a constant introduced for dimensional reasons, $\tilde{g}'$ and $\tilde{B}_{\nu}$ are a constant and a (gauge) field that will be illustrated to be related with the weak hyper-charge coupling constant and gauge field, respectively.\\
It is necessary to introduce matter fields and to extend the local Lorentz group to this scenario.\\
Spinor fields can be here generalized by appending a suitable functional dependence on the extra-coordinate and a normalization factor, i.e. ${^{5}\chi}_{i}(x^{\rho},x^{5})=\frac{1}{\sqrt{L}}{^{4}\psi}_{i}(x^{\rho})e^{\frac{i2\pi N_{i}x^{5}}{L}}$, where $L$ is the length of the extra-ring. If $\gamma^{5}$ is chosen as the fifth Dirac matrix, 5-D Dirac equations show a natural chirality for right- and left-handed spinors.\\ 
After the implementation of KK paradigm, the generators and the gauge fields of the 5-D Lorentz group are no longer given by $\Sigma^{\bar{A}\bar{B}}=\frac{i}{4}[\gamma^{\bar{A}},\gamma^{\bar{B}}]$; as suggested by the set of allowed KK transforations, 4-D generators are still valid, and $SU(2)$ weak isospin group can be looked for in the extra-D generators: from (\ref{conserved}), it follows that $J^{\bar{i}\bar{5}\mu}\equiv J^{{\mu i}}_{SM}\Rightarrow \Sigma^{\bar{i}\bar{5}}\propto -ig\frac{\sigma^{i}}{2}P_{L}$. Weak isospin gauge fields are identified with the pertinent extra-D gauge fields, while $A^{\bar{A}\bar{B}}_{5}$ must vanish in order to assure the correct world transformations for the 4-D fields.\\ 
Accordingly, left-handed doublets
\begin{equation}\nonumber
X_{f_{i}L} \equiv
\left(
\begin{array}{c}
\chi_{\nu_{i}L}\\
\chi_{l_{i}L}
\end{array}
\right)\equiv \frac{1}{\sqrt{L}}\Psi_{iL}e^{\frac{i2\pi}{L}N_{iL}x^{5}}:
\end{equation}
and right-handed singlets $\chi_{f_{i}R}$ are the natural spinor bases of this model.\\
Structure constants are given by the commutators of the generators: the only non-vanishing commutators are those defined in section 1 and $[\Sigma_{\bar{i}\bar{5}},\Sigma_{\bar{j}\bar{5}}]=C^{\bar{k}\bar{5}}_{\bar{i}\bar{5}\bar{j}\bar{5}}\Sigma_{\bar{k}\bar{5}}$; $\Sigma^{\bar{0}\bar{5}}$ is a \slshape redundant degree of freedom\upshape, and must be set equal to zero.
The commutators $[\Sigma_{\bar{a}\bar{b}},\Sigma_{\bar{i}\bar{5}}]=0$, which vanish because the generators act on different spaces, define the splitting of the 5-D Lorentz group after the implementation of KK paradigma, $SO(4,1)\rightarrow SO(3,1)\otimes SU(2)$, and will be the starting point for the introduction of Ashtekar formalism in this scheme.\\
Conserved quantities are the weak isospin charge, according to (\ref{conserved}), and the weak hyper-charge, whereas $SU(2)$ and $U(1)$ gauge transformations for spinors are obtained by the extra-D KK coordinate transformation. 
\section{Restoration of the Electro-Weak model}
Collecting all the terms together \cite{lem2006}, after dimensional reduction, the Lagrangian density consists of the Lagrangian density of the Electro-Weak model and that of the Lorentz gauge theory, as can be easily worked out by taking into account the properties of the connections and by eliminating the dependence on the extra-coordinate. As in usual KK theories, relations between the extra-components of the metric tensor and hyper-charge gauge objects can be found.
\section{Ashtekar's formalism?}
In 5 D, the total group we are dealing with is $SO(4,1)\rightarrow SO(3,1)\otimes SU(2)\rightarrow SU(2)\otimes SU(2)\otimes SU(2)$. An ADM splitting on the 5-D manifold gives a $4+1$ reduction: on the 4-D Euclidean resulting manifold, where the index $5$ plays a role analogous to that of $0$ in standard scenarios, a further splitting can be performed \cite{lam2006}. Here, $SO(4) \rightarrow SO(3) \otimes SU(2)$, and Ashtekar variables \cite{r91} can be defined: $C^{i}_{\alpha}=\epsilon^{i}_{jk}\omega^{jk}_{\alpha}$ for $SO(3)$, and $ {^{\pm}A}^{i}_{\alpha}={^{\pm}A}^{i5}_{\alpha}=\omega^{i5}_{\alpha}\pm\frac{1}{2}C^{i}_{\alpha}$ for $SU(2)$; here, $\omega^{jk}_{\alpha}$ denote the Euclidean spin connections, and $\epsilon^{ijk}\equiv\epsilon^{i5jk}$ (we recall that, in 5-D KK theory, the index $5$ is scalar). Bein vectors are evolutionary variables while connections are not: they can be mixed up in order to find suitable variables, but, as a perspective, a more general theory can be looked for, such that the $SU(2)\otimes SU(2)$ group can be built making use of 4-D objects only, thus recasting the original Ashtekar formalism, without mixing up ${^{+}A}^{i}_{\alpha}$ and $e^{i}_{\alpha}$.

\bibliographystyle{ws-procs975x65}
\bibliography{ws-pro-sample}

\end{document}